\documentclass{emulateapj}

\usepackage{color}
\usepackage{natbib}
\usepackage{ulem}
\usepackage{soul}
\usepackage{graphicx}

\def\cb279{3C~279}
\def\d1510{PKS~1510-089}
\def\thesource{{3C~454.3}}
\def\oj287{OJ~287}

\def\smarts{\textit{SMARTS}}
\def\fermi{\textit{Fermi}}
\def\pflux{ph s$^{-1}$ cm$^{-2}$}
\def\g{$\gamma$}

\begin{document}

\title{A Consolidated Framework of the Color Variability in Blazars: Long-Term Optical/Near-Infrared Observations of 3C 279}

\author{Jedidah C. Isler\altaffilmark{1, 2}, C.M. Urry\altaffilmark{3,4}, P. Coppi\altaffilmark{3}, C. Bailyn\altaffilmark{3,}, M. Brady\altaffilmark{3}, E. MacPherson\altaffilmark{3},  M. Buxton\altaffilmark{3}, 
I. Hasan\altaffilmark{5}
}

\altaffiltext{1}{Department of Physics and Astronomy, Vanderbilt University, 6301 Stevenson Center, Nashville, TN 37235; jedidah.isler@vanderbilt.edu}
\altaffiltext{2}{NSF Astronomy \& Astrophysics Postdoctoral Fellow}
\altaffiltext{3}{Department of Astronomy, Yale University, PO Box 208101, New Haven, CT 06520-8101}
\altaffiltext{4}{Department of Physics and Yale Center for Astronomy and Astrophysics, Yale University, PO Box 208121, New Haven, CT 06520-8120}
\altaffiltext{5}{Department of Physics, University of California, Davis, One Shields Avenue, Davis, CA 95616}


\begin{abstract}
We evaluate the optical/near-infrared (OIR) color variability of \cb279\ in both \g-ray flaring and non-flaring states over 7-year timescales using the Small and Medium Aperture Research Telescope System (\smarts) in Cerro Tololo, Chile and \g-ray fluxes obtained from the \fermi\ \g-ray Space Telescope. This observing strategy differs from previous blazar color variability studies in two key ways: 1) the reported color variability is assessed across optical through near-infrared wavelengths, and 2) the color variability is assessed over timescales significantly longer than an individual flare or ground-based observing season. We highlight \cb279\ because of its complex color variability, which is difficult to reconcile with the simple `redder when brighter' behavior often associated with Flat Spectrum Radio Quasar (FSRQ) color variability. We suggest that the observed OIR color changes depend on a combination of the jet and disk emission. We parametrize this behavior in terms of a single variable, $\zeta^m_n$, representing a smooth transition from disk-dominated, to a mixed contribution, to a jet-dominated system, which provides an explanation of the long-term OIR color variability in the same blazar over time. This suggests a general scheme that could apply to OIR color variability in other blazars.
\end{abstract}

\section{Introduction}
Blazars are radio-loud active galactic nuclei (AGN) whose relativistic jet is pointed at small angles  with respect to the line of sight \citep{antonucci93, Urry95}. The bulk relativistic outflow of these charged particles, along with the small viewing angles, induces Doppler beaming along the direction of the outflow such that the beamed emission dominates the bolometric luminosity. This makes blazars ideal laboratories for the study of relativistic jets.

Blazars can be divided into two sub-classes: Flat Spectrum Radio Quasars (FSRQs)  and BL Lac objects (BL Lacs). While FSRQs generally have prominent emission lines, BL Lacs are characterized by the absence or extreme weakness of emission lines, with equivalent width $<$ 5\AA\ \citep{Urry95}. Generally, FSRQs have higher jet luminosity and lower peak synchrotron frequencies in the broadband spectral energy distribution (SED), and BL Lacs have lower jet luminosity and higher synchrotron peak frequencies \citep{Fossati98}, although as observations of blazars have probed a wider range of \g-ray and radio luminosities, it appears that there is likely a blazar envelope rather than a sequence \citep{Meyer12b}.  When characterized by their peak synchrotron frequencies, blazars can be subdivided as low- ($\lesssim$~10$^{14}$~Hz), intermediate- (10$^{14-15}$~Hz) and high- synchrotron peaked sources ($\gtrsim$~10$^{15}$~Hz) \citep{Padovani96, Abdo10}.

The location of the peak synchrotron frequency, $\nu^{peak}_{syn}$, is determined in part by the power-law energy density distribution of non-thermal electrons producing some of the observed optical/near-infrared (OIR) emission. Blazars can also have strong thermal components (the accretion disk and/or broad emission line region) that contribute significantly in the OIR.

The optical color variability of blazars has historically been associated with blazar type: FSRQs have been observed to be redder when brighter \citep{Ramirez04,Gu06, Osterman08, Osterman09, Bonning12}, likely due to an increase in the relative contribution of the beamed jet emission to the less variable, bluer accretion disk emission. Alternatively, BL Lacs are often observed in bluer when brighter states, likely a result of significantly less radiative cooling of highly accelerated electrons off an under-luminous or non-existent accretion disk \citep{Massaro98, Villata02, Vagnetti03, Gu11b, Gaur12}. These observations have been extrapolated to represent the ``typical'' behavior of FSRQs and BL Lacs. 

Many departures from this putative dichotomy in color variability have been reported with some FSRQs becoming bluer when brighter \citep{Gu11}, and at other times, redder when brighter \citep{Wu11}; some blazars remain achromatic with increasing brightness \citep{Ghosh00, Stalin06}. \citet{Ghisellini97} observed BL Lac S5~0716+714 for 5 months using the \textit{International Ultraviolet Explorer}, with two weeks of overlapping \textit{EGRET} observations. Observing in the optical bands, they find that color flattens with increasing brightness during the monitored \g-ray flares, but that this color-brightness relationship is insensitive to the full 5 month trends. 
 
\citet{Gu06} found that of 8 red blazars observed over 5 months, 5 BL Lacs were bluer when brighter, 2 FSRQs were redder when brighter, and one FSRQ was achromatic over the observation period. \citet{Bottcher05} observed BL Lac 3C~66A over 9 months in 2003, with a core campaign of 3 months, and reported a change in the optical color variability from redder when brighter to a flattening at the brightest magnitudes.

However, these color variability studies were limited to short timescale observations (from a few days to a few months) of flaring episodes in specific blazars, and often only in the optical bands. Given the phenomenological differences between BL Lacs and FSRQs, when they are preferentially observed during \g-ray flares, it is possible to incorrectly infer a dichotomy of behavior when a continuous transition might exist. Such behavior may only be detected when the source is monitored over long time frames to allow for a larger diversity of source behaviors. 

As part of our Small and Medium Aperture Research Telescope System (\smarts) OIR monitoring of blazars \citep{Bonning12}, we observed detailed color variations in both BL Lac objects and FSRQs, across a large variation in brightness by analyzing 3 years of OIR color variability in 12 \fermi-bright blazars. This study also concluded that the fractional variability amplitude increases blueward of the peak synchrotron frequency and that, for many FSRQs, the infrared is more variable than the optical during jet flares, suggesting an increase in the presence of non-thermal emission in the infrared bands.

In the current work, we investigate the OIR color variability of the FSRQ \cb279\ (z~=~0.536) and extend previous color variability analyses both temporally, with on-average nightly cadence over 7 years, and spectrally by including near-infrared photometry, which greatly extends the color baseline, to investigate the long-term OIR color variability. 

\cb279\ has a well-studied broadband variability profile during both \g-ray active and quiescent states \citep[e.g.][]{Maraschi94, Hartman96}. \citet{Wehrle98} found that the broadband spectral energy distribution showed inverted spectral variations such that the optical emission was lower by a factor of 2 when the \g-ray emission was higher by a factor of 4 during an active phase. Several studies have found evidence for correlated variability between the optical and \g-ray energies, although correlation with IR and \g-rays has not been as consistently detected, even in the \fermi\ era \citep{Bonning12}. Thus, its complex spectral behavior makes it an ideal candidate to investigate OIR color variability.  

We present the details of the \smarts\ optical, near-infrared and \fermi\ \g-ray monitoring program, data reduction and analysis in \S~\ref{sec:photmeth}. The multiwavelength light curve and OIR color variability of \cb279\ on short- and long-timescales are presented in \S~\ref{sec:photres}. In \S~\ref{sec:photmodel}, we propose a consolidated framework for understanding the OIR color variability in blazars in terms of the slope of the OIR color variability diagram, and then discuss the significance of our current findings with respect to the apparent blazar-color variability dichotomy in \S~\ref{sec:photdisc}. We summarize our conclusions in \S~\ref{sec:photconc}.

\begin{figure*}
\includegraphics[width=1\linewidth]{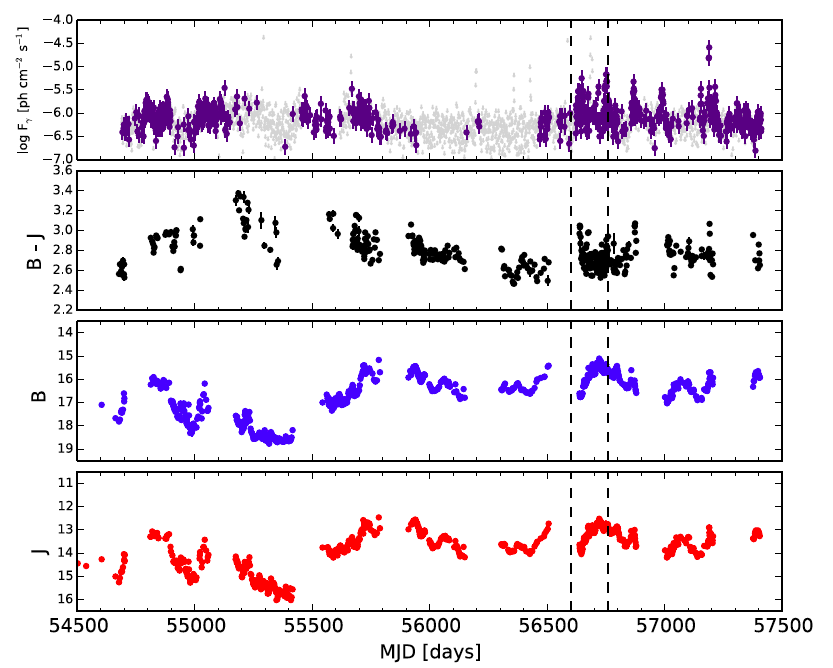}
\caption{The \fermi\ [E$>$100 MeV] \g-ray (top panel) and \smarts\ B~$-$~J color (second panel), B-band and J-band light curves for \cb279\ over 7 years of \smarts\ monitoring. \smarts\ data are given in magnitudes and \fermi\ photon fluxes are given in units of log \pflux. The \g-ray fluxes with TS $>$16 are plotted in purple points and upper limits are represented by grey downward facing arrows. Error bars are included in all panels, but for the \smarts\ data, uncertainties are approximately the same size as the symbols. Vertical dashed lines mark the location of the \g-ray flare identified for short-term color variability in Figure~\ref{fig:last100}. The total flux and B~-~J color vary over the course of these multi-year observations. \label{fig:lc3c279}}
\end{figure*}

\section{Observations \& Data Analysis}\label{sec:photmeth}
Optical/near-infrared photometry of \cb279\ between 2008 May 17 and 2015 December 31 was obtained using the \smarts\ 1.3m~+~ANDICAM in Cerro Tololo, Chile. ANDICAM is a dual-channel imager with a dichroic that simultaneously illuminates an optical CCD and IR imager yielding photometry from 0.4 - 2.2~$\mu$m (spanning the Johnson-Cousins BVRJK bands). Photometric cadence is roughly nightly as long as the source is visible from Cerro Tololo and as weather permits.

The OIR photometric reduction and analysis were described in \citet{Bonning12} and will only be summarized briefly here.  Optical comparison stars were calibrated from Landolt standards on photometric nights. The near-infrared magnitudes are calibrated using Two Micron All Sky Survey magnitudes \citep{Skrutskie06} of a secondary star in the same field as \cb279. In the optical bands, the photometric uncertainties are dominated by the random errors in the photometric calibration of the comparison star magnitude from the USNO-B catalog, or $\sigma_{opt}\simeq$~0.03 magnitudes \citep{Monet03}. We estimate the uncertainty in the near-infrared bands to be  $\sigma_{IR}\simeq$~0.05 magnitudes.

The \smarts\ B- and J-band, B~$-$~J color and \fermi\ \g-ray flux light curves for \cb279\ are shown in Figure~\ref{fig:lc3c279}. We refer the reader to the \smarts\ website\footnotemark[1], where OIR photometry is publicly available for the full \smarts\ source list. Data included in the present analysis have had errant data points that are associated with instrumental aberrations and other non-physical origins removed. These points are easily identified by single-band deviations in the \smarts\ data \sout{and datapoints} that vary by more than 3 standard deviations from nearby measurements. \footnotetext[1]{http://www.astro.yale.edu/smarts/glast/home.php}

\fermi/LAT data were obtained from the \fermi\ Science Support Center website\footnotemark[2] between 2008~$-$~2015. 
Pass 7 reprocessed data (event class 3) were analyzed using \fermi\ Science Tools (v9r33p0) with customized scripts to automate the likelihood analysis.
Galactic diffuse model (gll\_iem\_v05\_rev1), isotropic diffuse background (iso\_clean\_v05) and instrument response functions (P7REP\_CLEAN\_V15) were utilized in the analysis. 
Data were constrained to time periods where the zenith angle was less than 100$^\circ$ to avoid Earth limb contamination and only photons within a 10$^\circ$ region centered on the source of interest were analyzed.  
The \g-ray spectra was modeled as a log-parabola, according to the spectrum type listed in the 3FGL catalog, with the photon flux and spectral index as free parameters. 
 \fermi\ light curves (E $>$ 100MeV) were calculated in one day bins to match the average \smarts\ cadence. 
Daily points for which TS $>$ 16 are plotted, where TS is the Fermi test statistic and $\sqrt{TS}$ is roughly equivalent to the source detection significance per integrated bin \citep{Mattox96,Abdo09}. 
\footnotetext[2]{http://fermi.gsfc.nasa.gov/ssc/data/access}

\section{Results}\label{sec:photres}
\cb279\ was regularly detected by \fermi\ with high \g-ray significance from 2008 November 30 through 2011 August 27 (MJD 54800 $-$ 55800), after which time the source underwent a period of relative jet quiescence. Subsequently, a \g-ray flare was detected between 2013 November 04 and 2014 April 13 (MJD 56600 $-$ 56760) with peak F$_\gamma$[E$>$100 MeV] $\simeq$ 7 $\times$ 10$^{-6}$ \pflux, and a corresponding OIR  flare with peak magnitudes B~$\simeq$~15.5 mag and J~$\simeq$~12.5 mag, as seen in Figure~\ref{fig:lc3c279}. 

\begin{figure}
\includegraphics[width=1\linewidth]{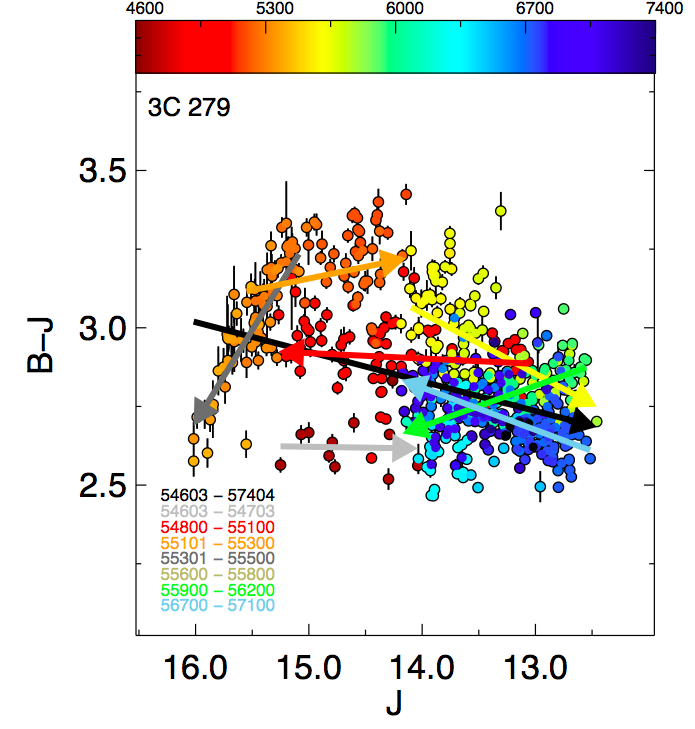}\caption{The \smarts\ OIR color variability diagram for \cb279\ from 2008~$-$~2015. The color bar indicates the progression of time in units of MJD$-$50000 and arrows indicate the linear least squares fits to the epochs in corresponding colors in the legend and in Table~\ref{tab:zeta}. The color of the arrows corresponds to approximately the epoch to which the linear fit was calculate. While the average OIR slope is bluer when brighter ($\zeta^-_- >$~0; black arrow) over the full \smarts\ observation period, other color variability responses can be seen on timescales of 100s of days.
The light grey is for achromatic color variability with $\zeta^0_- \simeq$~0 at MJD 55101$-$55300 and the orange arrow is for redder when brighter with $\zeta^+_- <$~0 at MJD 54603$-$54703. Details about the remaining color variability can be found in the text. \label{fig:colmag}}
\end{figure}

\begin{figure}
\includegraphics[width=1\linewidth]{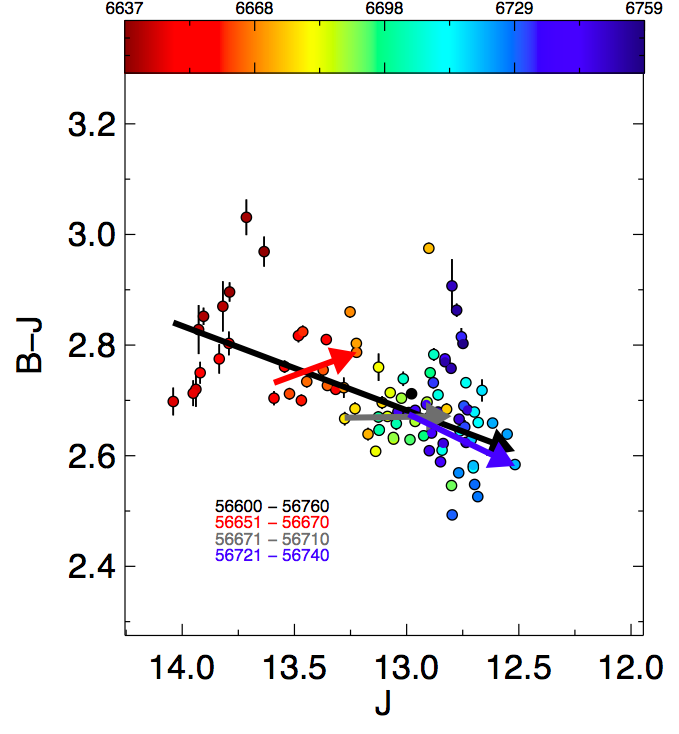}
\caption{The \smarts\ OIR color variability diagram for \cb279\ from 2013 November 4 $-$ 2014 April 13 (MJD 56600 $-$ 56760). The arrows denote the least squares fit to the 160$-$day flaring period in subsets of tens of days and the MJDs listed correspond to arrows of the same color. The arrow color indicates the behavior of the OIR color variability, as described in Figure~\ref{fig:model}. The black line denotes $\langle \zeta^m_n \rangle$ over the flare. The OIR color variability can again be observed, this time on timescales of 10s of days, although the phenomena on short timescales and long timescales (as shown in Figure~\ref{fig:colmag}) have different physical drivers, as described in the text. Between MJD 56651$-$56670, \cb279\ is redder when brighter ($\zeta^+_- <$~0; red arrow), but as the flare progresses, $\zeta^0_- \simeq$~0 flattens (grey arrow) until the color again becomes blue with increasing brightness ($\zeta^-_- >$~0; blue arrow) during MJD 56721$-$56740.   \label{fig:last100}}
\end{figure}

We plot the OIR color variability diagrams, i.e. B~$-$~J color versus J-band magnitude, for \cb279\ in Figures~\ref{fig:colmag} and \ref{fig:last100}. Figure~\ref{fig:colmag} shows the OIR color variability over the full \smarts\ monitoring period between 2008~$-$~2015. Figure~\ref{fig:last100} illustrates the OIR color variability for the 2013 November 04 through 2014 April 13 (MJD 56600 $-$ 56760) \g-ray flare lasting 160 days. 

For the purposes of this work, ``long-term" indicates the 7 years of data collected for \cb279, which includes several jet activity states, that is, enough time for \cb279\ to transition from \g-ray quiescence to activity more than once. We also parse the data into smaller epochs to characterize short-term variability patterns. While this separation of the data is, by definition, arbitrary, it has been reported that measuring blazar variability on logarithmic timescales allows for the most complete and least biased characterization of the data \citep{Chatterjee12}.

We measure the slope of the OIR color variability of a given flare with a $\Delta$t of order a few days, 10s of days, 100s of days and 1000s of days. We fit a linear, least squares regression to the data. 

To quantify the OIR color variability, we measure the slope of the B~$-$~J versus J-band magnitude, as plotted in  Figures~\ref{fig:colmag} $-$  \ref{fig:model} and defined as:

\begin{equation}
\zeta^m_n \equiv \frac{\Delta(B - J)}{\Delta J},
\end{equation} 

where m is the sign of $\Delta$(B~$-$~J) and n is the sign of $\Delta$J along the time sequence. Note that there are two physically distinct ways for $\zeta^m_n <$~0 and $\zeta^m_n >$~0 to result. If the sign of $\Delta$(B~$-$~J) and $\Delta$J are the same, $\zeta^m_n$~$>$~0. If the sign of $\Delta$(B~$-$~J) and $\Delta$J are different, $\zeta^m_n$~$<$~0. The epochs of OIR color variability analyzed in this work are listed in Table~\ref{tab:zeta}.

In Figure~\ref{fig:colmag}, the average slope of OIR color variability, \textbf{$\langle \zeta^m_n \rangle$}, over the 7 years is indicated by the black arrow;  \cb279\ is bluer when brighter ($\zeta^-_- >$~0) over the longest timescales observed by \smarts. However, it can also be clearly seen that the OIR color variability deviates significantly from this trend on shorter timescales. For example, redder when brighter ($\zeta^+_- <$~0; orange arrow), redder when fainter ($\zeta^+_+ >$~0; blue arrow), achromatic ($\zeta^0_- \simeq$~0; light grey arrow), bluer when brighter ($\zeta^-_- >$~0 yellow arrow) and bluer when fainter ($\zeta^-_+ <$~0; dark grey and green arrows) color variability can also be seen.

In addition, the average OIR color $\langle B~-~J \rangle$ $\gtrsim$~2.5 over $\vert\Delta$J$\vert$~=~3 mag; while there is clear color variability present, there may be a lower limit to how blue the OIR color can be. For \cb279, this is likely due to the spectrum of the underlying, but relatively luminous accretion disk \citep{Pian99}.

We report similar OIR color variability on short timescales, as seen in Figure~\ref{fig:last100}. The turnover in $\zeta^m_n$ is evident from redder when brighter ($\zeta^+_-<$~0; red arrow) to nearly flat ($\zeta^0_- \simeq $~0; grey arrow) to bluer when brighter ($\zeta^-_->$~0; blue arrow) over the 160 day flare. We note that the entire flare plotted in Figure~\ref{fig:last100} took place when \cb279\ was brighter, on average, than the previous 6 years of monitoring ($\langle$J$\rangle$~$\sim$~13.09 mag) and that the average OIR color is bluer when brighter ($\zeta^-_- >$~0; black arrow) during the flare. Thus, while the overall color variability is bluer as the source brightens over the \g-ray flare, there are changes in the color variability on timescales of tens and hundreds of days.

\begin{deluxetable}{cccc|ccc}
\tablecaption{OIR Color Variability Epochs and Fit Parameters  \label{tab:zeta}}
\tablehead{\colhead{Start} & \colhead{End} & \colhead{$\Delta$t} & \colhead{$\langle$ J $\rangle$} & \colhead{$\zeta$} & \colhead{$\zeta^m_n$} & \colhead{CV}\\ 
 \colhead{(MJD)} & \colhead{(MJD)} & \colhead{(days)} & \colhead{(mag)} & \colhead{} & \colhead{} & \colhead{}}
\startdata
    54603 & 57404 & 2801 & 13.83 & 0.094 & \textbf{$\zeta^-_-$} & BWB \\\hline
   54800 &        55100  & 300&14.29 & 0.017 & $\zeta^0_+$ & ACH \\
  	54603 & 54703 & 100 & 14.65 & 0.006 &$\zeta^0_-$ &ACH \\
  	 55101 & 55300 & 199 & 14.91& -0.075 & $\zeta^+_-$ & RWB\\
  	55301 &        55500 & 199&15.63 & -0.587 &$\zeta^-_+$ & BWF\\
 	55600 &        55800 & 200&13.42 & 0.193 &$\zeta^-_-$ & BWB\\
 	55900 &       56200 & 300&13.25 & -0.129 &$\zeta^-_+$  & BWF\\
  56700 & 57100 & 400 & 13.21 & 	0.134 &$\zeta^+_+$ & RWF\\ \hline
  56600 & 56760 & 160 & 13.09 & 0.153 &$\zeta^-_-$ & BWB \\
   56600 &56650 & 50&13.86 & 	-0.750 &$\zeta^-_+$ & BWF\\
   56651 & 56670 & 19&13.41 & -0.150 &$\zeta^+_-$ &RWB \\
	56671 & 56710 & 39&13.02 & -0.006 &$\zeta^0_-$ &ACH \\
	 56711 & 56720& 9&12.70 & -0.188  &$\zeta^-_+$ &BWF \\
	 56721 & 56740 & 19 &12.76 & 0.20 &$\zeta^-_-$ & BWB \\
	 56741 & 56760 & 19&12.82 & -0.386 &$\zeta^+_-$ & RWB
\enddata
\tablecomments{Column contents: (1) and (2) Start and End Dates in MJD that define each epoch; (3) $\Delta$t $-$ the number of days over which $\zeta^m_n$ is calculated, (4) $\langle$J$\rangle$ $-$ the average J-band magnitude for each epoch, (5) $\zeta$ $-$ the calculated value of the OIR slope, (6) $\zeta^m_n$ $-$ as defined in Equation 1 with m,n specified and (7) CV $-$ the OIR color variability as indicated by the arrows in Figure~\ref{fig:model}. Abbreviations for CV: BWB $-$ bluer when brighter; ACH $-$ achromatic; RWB $-$ redder when brighter; BWF $-$ bluer when fainter; RWF $-$ redder when fainter.}
\end{deluxetable}

\section{Long-term OIR Color Variability:\\ A Continuum, not a Dichotomy}\label{sec:photmodel}
We suggest a simple framework to describe the short- and long-term OIR color variability observed in blazars. Specifically, these data can be explained, in many cases, by the differing relative contribution of the thermal disk and non-thermal jet emission. 

Figure~\ref{fig:model} is a schematic representation of the OIR color variability in a typical blazar. The shaded regions represent distinct physical regimes: 1) accretion disk dominant color, 2) an admixture of disk-jet colors, and 3) jet dominant color. Offset from each OIR color ``state" is a representative SED to help the reader interpret the spectral shape of an OIR color variability change, but does not indicate the SED of a specific blazar.

The observed color variations are described by a change in the jet emission, disk emission, or a combination of the two. Thus, while the SED shape is expected to differ from source to source as it depends on the source jet and disk luminosity, the relative positions of the states with respect to one another are physically robust.

We use empirical data from \cb279, as shown in Figures~\ref{fig:colmag} and \ref{fig:last100}, to outline this framework, which explains how blazars can smoothly evolve from a jet-quiescent, disk-dominated OIR color profile to an actively jet flaring state and associated OIR color profile and back to jet-quiescent state.  

\subsection{Accretion Disk Dominant Color} \label{sec:quiet} 
When the blazar jet is quiescent, the accretion disk spectrum is often observable in the OIR-UV regime \citep{Pian05}. It is believed that the temperature of the accretion disk scales with black hole mass \citep{Shakura73, Frank02} and for black hole masses M$_{BH}$ $\gtrsim$ 10$^8$M$_\odot$ the peak disk luminosity is in the extreme ultraviolet; for \cb279, T$_{disk}$~$\sim$~20~000~K \citep{Pian99}. It follows that in the OIR regime, the thermal disk spectrum is bluer, since it is on the exponentially increasing side of a multi-temperature blackbody profile with vanishingly little contribution to the IR (or redder) bands \citep{Ghisellini13b}. The OIR color would therefore be bluer, as represented schematically by the blue ellipse in lower left of Figure~\ref{fig:model}. This state can be seen in the OIR color variability of \cb279\ as shown in Figure~\ref{fig:colmag} where J $\simeq$ 16 mag and  B~$-$~J $\simeq$ 2.6.

Any increase in jet activity due to an increase in synchrotron flux would cause a reddening of the OIR color and a decrease in the magnitude  of the source (i.e. brightening), due to the differing spectral shapes as shown in Figure~\ref{fig:model} and described in Section~\ref{sec:admixture}. This would result in $\zeta^m_n<$~0, which is represented by the red ($\zeta^+_-$) and blue ($\zeta^-_+$) arrows in Figure~\ref{fig:model}, and represents the color evolution as the synchrotron flux increases and decreases, respectively. During jet quiescent states, we expect the OIR color to be relatively blue and the intensity to be relatively faint. In addition, the OIR color should be well-localized in this parameter space since the blazar accretion disk luminosity does not vary significantly on short timescales \citep{Lira11}. 

\subsection{Admixture of Accretion Disk-Jet Color}\label{sec:admixture} 
At the onset of a jet flare, the non-thermal synchrotron radiation will increase from the radio through the OIR, UV and/or soft X-rays depending on $\nu^{peak}_{syn}$. This increase in non-thermal emission results in an increased number of non-thermal photons with a redder spectral shape, thus moving simultaneously up and to the right ($\zeta^+_-<$~0) as indicated by the red arrow in Figure~\ref{fig:model}.

For \cb279, which is an LSP FSRQ, the highly variable non-thermal jet emission is proportionally redder than the thermal disk emission in the OIR band peaking in the infrared because of the location of the synchrotron peak, so as the jet flares, the OIR color becomes redder when brighter ($\zeta^+_-<$~0). This color evolution will persist as long as the observed jet luminosity is below the disk luminosity. 

However, as the jet emission increases and becomes more comparable to the disk emission, $\zeta^0_- \rightarrow$~0 with increasing brightness due to the differing spectral shape of the two components. Namely, the proportionally redder spectrum of the jet begins to rival and ultimately overcome the bluer disk emission due to Doppler beaming. In that case, the observed OIR color variability flattens before either turning blue with increasing jet contribution ($\zeta^-_- >$~0; as described in the next section) or fading back into a jet-quiescent phase ($\zeta^-_+ <$~0). Sources with this combination of disk and jet emission will occupy the purple rectangular region representing approximately equal contributions in color-magnitude space of both thermal and non-thermal emission; such color ``saturation", $\zeta^0_- \rightarrow$~0, has been observed in several blazars \citep{Villata06, Bonning12, Chatterjee13}. This is most clearly seen in Figure~\ref{fig:colmag} between MJD 54603~$-$~54703, where B~$-$~J is relatively constant; the fit parameters for this epoch can be seen in Table~\ref{tab:zeta}.

\subsection{Jet Dominant Color} 
Since the non-thermal jet emission is Doppler boosted and the disk emission is not, the observed jet luminosity will completely swamp the disk luminosity even if the intrinsic luminosities are similar. Once the jet emission has swamped the disk emission in the OIR regime, the OIR color is determined by the individual relativistic plasma ejections causing the flaring. We observe two distinct OIR color variability behaviors in the jet dominated state: i) bluer when brighter ($\zeta^-_- >$~0) and, ii) redder when brighter ($\zeta^+_- <$~0). In the former case, while the \g-ray flare is dominant, the OIR color becomes bluer when brighter, likely due the electrons producing the synchrotron emission reaching higher energies before radiatively cooling on ambient photons from the disk, broad line region and/or molecular torus (depending on distance from central source). Thus, the bluer when brighter behavior is related to the spectral evolution of the non-thermal emission to higher intensities \textit{and} higher spectral frequency. This is the OIR color variability behavior expected in a blazar jet dominant state and is indicated by the blue arrow ($\zeta^-_- >$~0) in Figure~\ref{fig:model}.

In the latter case, the additional \g-ray flares may not reach as high an energy as the accelerated electrons that were previously produced, and thus appear redder even as the source continues to brighten. This reddening is physically distinct from (although phenomenologically similar to) the redder when brighter behavior that is observed during the admixture of disk and jet emission.  The jet-dominated increase in redness is due to distinct electron acceleration episodes and \textit{not} to the relative global colors of the disk and jet emission. Therefore, this is an intrinsically transient state that persists only as long as a detectable difference in particle acceleration and resultant radiation spectrum is present. This transient state is governed by the particle cooling time. 

Both cases of jet-dominated behavior have been observed in \cb279\ on short timescales, as can be seen in Figure~\ref{fig:last100}, namely bluer when brighter, $\zeta^-_- >$~0 (blue arrow; MJD 56721 $-$ 56740)  and redder when brighter $\zeta^+_- <$~0 (red arrow; MJD 56651 $-$ 56670) behavior seen on timescales of tens of days. The fits to these epochs can be found in Table~\ref{tab:zeta}.

\subsection{Cooling Mechanisms}
\begin{figure*}
\includegraphics[width=1\linewidth]{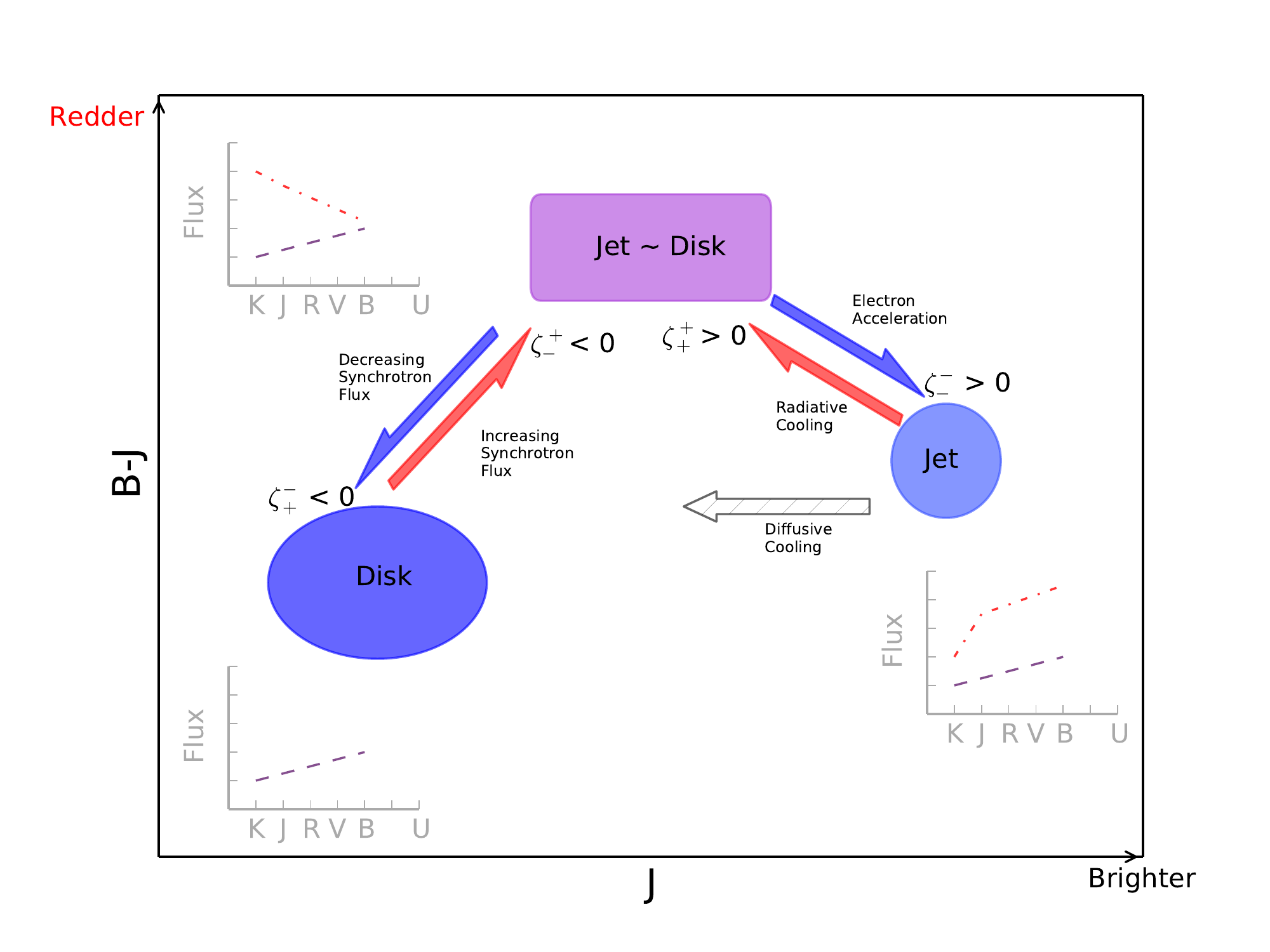}
\caption{A schematic representation of the OIR color variability diagram (B~$-$~J color versus the J-band magnitude) for a given blazar on long timescales. B~$-$~J becomes increasingly redder with increasing y-value, and J-band magnitude decreases (gets brighter) to the right. The accretion disk is inherently bluer and relatively faint compared to jet flaring states; the jet is inherently redder (in the OIR bands), and in particular as the jet luminosity begins to rival and swamp the disk luminosity. As the disk gets brighter, it will tend toward a bluer OIR color. When the jet brightens relative to the disk, it will become redder ($\zeta^+_-<$~0), while bluer colors result from the jet diminishing in brightness relative to the disk ($\zeta^-_+<$~0). Once the jet luminosity completely swamps the disk, further jet brightening will result in bluer colors ($\zeta^-_->$~0),  due to particle acceleration, and will fade toward the red ($\zeta^+_+>$~0), due to radiative losses. SEDs are inlaid to approximate luminosity contribution and spectral shape of the disk (\textit{purple dashed}) and jet (\textit{red dot-dashed}) emission and are meant to illustrate the three ``states" described in the text; namely disk-dominant colors (\textit{dark blue ellipse}), and admixture of disk-jet colors (\textit{purple rectangle}) and, jet-dominant colors (\textit{light blue circle}). Colored arrows denote the transitions between ``states" and the arrow color indicates the OIR color achieved during the transition. Diffusive cooling is shown as an arrow with hatches since it is believed to be an achromatic transition. The transitions occur relative to each other asdepicted in the diagram, thus while the location of the ``state" changes in this parameter space is expected to differ from source to source, any blazar can occupy any region of this diagram, moderated by degree of jet activity (or quiescence). \label{fig:model}}
\end{figure*}

As the \g-ray flare dissipates, the cooling mechanisms are also imprinted on the OIR color evolution, as described below and shown in Figure~\ref{fig:model}.

\begin{itemize}
\item \textit{Radiative Cooling:} The highest energy electrons in the jet dominant state produce a bluer when brighter spectrum as they are accelerated ($\zeta^-_- >$~0), but as that radiation dissipates, the electrons cool by reddening (Figure~\ref{fig:model}, red arrow; $\zeta^+_+ >$~0). However, as the total flare dissipates and again becomes comparable to the disk luminosity, the OIR color variability will decrease toward the blue (Figure~\ref{fig:model},  blue arrow; $\zeta^-_+ <$~0). This type of radiative cooling can be seen by the blue arrow in Figure~\ref{fig:colmag}, near J~$\simeq$~13.5 mag, where the source is becoming redder as it is fading in brightness ($\zeta^+_+>$~0).
\item \textit{Diffusive (or Escape) Cooling:} If the highest energy electrons escape the emitting region on timescales much shorter than the radiative cooling timescale, these accelerated electrons will not contribute to the OIR color evolution even as the energy density decreases, causing the flare to fade approximately achromatically, especially as the source fades in luminosity in the jet-disk to disk states. Since the cooling rate is energy dependent, the higher energy end of the electron energy density distribution is evacuated faster than the lower energy end and the system cools with virtually no change in color. Such achromatic cooling is indicated by the red arrow in Figure~\ref{fig:colmag} with magnitude 15 $<$ J $<$ 14 ($\zeta^0_+$~$\simeq$~0).
\end{itemize}

\section{Discussion}\label{sec:photdisc}
\subsection{Previous Color Variability Studies}
Previous blazar color variability studies found in the literature find two distinct responses: 1) BL Lacs become bluer when brighter and, 2) FSRQs become redder when brighter \citep[e.g.][]{Massaro98, Ghosh00, Trevese02, Villata02, Damicis02, Vagnetti03, Gu06, Ramirez04, Wu11}. These results often led to the interpretation that these are the only optical color variability behaviors that could arise. The bluer when brighter behavior observed in BL Lacs was thought to be due to electrons being accelerated to preferentially higher energies before radiatively cooling, and FSRQs were thought to become redder when brighter because of the addition of redder, non-thermal jet emission to an already blue, thermal disk component. These interpretations are aligned with the framework we present here, but we suggest that the apparent dichotomy in color variability of FSRQs and BL Lacs arises because they are preferentially observed while flaring, causing a selection effect such that the two extreme ends of the color variability continuum are most regularly observed. Thus, the previous color variability results can be thought of as `snapshots' of the long-term OIR color variability framework presented here, as demonstrated using the blazar \cb279.

\citet{Ruan14} point out that optical color variations could be the result of ``hot spots'' in the accretion disk emission. Our present analysis cannot resolve changes in disk luminosity at this scale and thus cannot be directly compared to the current study.

\subsection{Estimating the Disk Luminosity using OIR Color Variability}
While blazars are generally jet-dominated, during phases of jet quiescence the thermal disk contribution can sometimes be seen. If one assumes that the disk does not vary as quickly as the jet, then on short timescales, one can measure the increasing non-thermal OIR contribution by holding the thermal contribution constant and measuring the difference in color. Thus, we can in principle determine solely from the OIR color (while confirming the presence of jet activity using simultaneous \g-ray flux measurements) whether we are in a disk-dominated, a jet-dominated, or an admixture of disk-jet regime. 

For sources that have highly luminous accretion disks, as the jet increases in luminosity and color contribution, the color variability will tend to flatten (less change in OIR color with increase in brightness) even as the magnitude continues to increase because the non-thermal --and beamed -- jet emission overtakes the thermal disk emission in those bands. More specifically, the competition between the differing spectral shapes of the thermal and non-thermal components will tend to produce a flat spectral shape in the OIR wavebands. This flattening in color variability space will persist until the jet has completely swamped the disk in both luminosity and color after which point, if the jet continues to intensify, the color variability would become bluer when brighter ($\zeta^-_- >$~0). Such flattening in the color variability has been observed in \cb279, as well as in \d1510\ and \thesource\ \citep{Bonning12}. This flattening can also account for a decrease in rate of color variability change observed in previous short-term analyses, which do not manifest in longer term data on the same source \citep{Ghisellini97, Bottcher05}.

\subsection{Extending the OIR Color Variability Diagram to Other Sources}
While we leave a fuller treatment of the OIR color variability of the entire \smarts\ sample to the next paper in this series, we do consider examples of divergent OIR color variability extant in the literature.
\citet{Bonning12} reported peculiar OIR color variability in low-synchrotron peaked BL Lac object \oj287\ over long timescales, where it was found to deviate from the bluer when brighter behavior generally associated with BL Lacs. While the cause of the complex OIR color variability in \oj287\ is still unclear, \citet{Bonning12} reported a number of physical scenarios that could give rise to such behavior. Among them: rapid and recurrent electron injections and an interplay between the accretion disk and jet emission; both scenarios fit within the framework presented here. 

Furthermore, it is important to note that diffusive/escape cooling is not the only way that a blazar could undergo minimal color variations while changes in observed magnitude are detected. If, for example, there are changes in the jet direction \citep[for early work on jet precession see][]{Gower82}, and hence changes in the Doppler boosting factor then the blazar magnitude could increase or decrease without a detectable change to the OIR color. This is true as long as the non-thermal spectrum maintains a constant slope throughout the wavelengths boosted in the observed OIR waveband due to precession; otherwise, a bluer OIR spectrum would result from an increase in Doppler boosting (and photon energy). We note that such jet precession has been postulated for \cb279\ \citep{Abraham98} and other blazars \citep{Caproni13, Qian14}, and cannot be ruled out with the present dataset.

The particle acceleration and cooling mechanisms that govern the OIR color variability presented here are physically motivated and present in every blazar system, so that although the specific color of a source during quiescent and flaring states will vary, we posit that the overall color variability evolution is robust. We note that the progression in color variability space does not always cover the full parameter space represented in Figure~\ref{fig:model}. For example, if the color from a jet flare does not rival or overtake the disk color, only the redder when brighter ($\zeta^+_- <$~0) to bluer when fainter ($\zeta^-_+<$~0) evolution will be observed. This can explain why FSRQs, which have very luminous disks, are often found on the redder when brighter ($\zeta^+_- <$~0) portion of the diagram and BL Lacs, which often have under-luminous or non-existent accretion disks, are often found on the bluer when brighter ($\zeta^-_- >$~0) portion of the diagram. 

A prediction of this framework is that high-synchrotron peaked blazars will not show redder when brighter ($\zeta^+_- <$~0) OIR color variability due to variability in the disk to jet luminosity because: a) their accretion disk is extremely weak or non-existent thereby removing the competing thermal emission in this regime and b) because the peak synchrotron frequency is so high, the OIR spectrum is bluer due to the spectral shape of the non-thermal emission and not due to the thermal disk emission as described in the model presented here. Low-synchrotron peaked blazars would be more likely to have a redder when brighter response because the range over which their peak synchrotron frequencies have been observed allow for such a behavior; the latter remains true for FSRQs. 

\subsection{OIR Color Variability and Optical Linear Polarization in \cb279}\label{pol}
Several studies on the optical polarization of individual blazars, as well as statistical studies of larger blazar samples have been undertaken in recent years \citep[see][for a review]{Hovatta17}, including for \cb279\ \citep{Larionov08, Abdo10c, Zhang15}. \citet{Kiehlmann16} report two electric vector position angle (EVPA) rotations in \cb279: 352$\degr$ during a \g-ray quiescent epoch (MJD 55100 $-$ 55310; epoch 1), and -494$\degr$ during a \g-ray flaring epoch (MJD 55658 $-$ 55850; epoch 2). They suggest that two distinct types of processes produce the optical polarization in \cb279: stochastic (during the low \g-ray brightness state) and deterministic (during the jet flaring state), based on the nature of the polarization variation. 

We have reported the OIR color variability of these regions above (see Table~\ref{tab:zeta}) and find that during the \g-ray quiescent period (epoch 1) $\zeta^+_- <$~0, indicating a transition from the disk-dominant phase with increasing contribution from the non-thermal (and polarized) jet emission. During the \g-ray flare (epoch 2), we report that $\zeta^-_- >$~0 which corresponds to a jet-dominant phase, where the individual flares dominate the OIR emission. Thus, our results are consistent with the \citet{Kiehlmann16} interpretation of the optical polarization results for these two epochs, such that the OIR color variability also indicate the degree to which this non-thermal and polarized emission is present in the optical bands. We note, however, that while it is not clear that optical polarization can be consistently correlated with \g-ray flaring, we suggest that in \cb279, all phases of the jet evolution should be detectable in the OIR color variability diagram.

\section{Summary and Conclusions}\label{sec:photconc}
The analysis of the optical/near-infrared color variability on FSRQ \cb279\ over long and short-timescales can be interpreted in a framework in which a relatively blue accretion disk and redder jet change in relative intensity. In addition, the jet color can change on a rapid timescales while the disk color changes more slowly. We find that the OIR color variability behavior of \cb279\  is alternatively redder when brighter, $\zeta^+_- <$~0, bluer when brighter, $\zeta^-_- >$~0, or achromatic, $\zeta^0_- \simeq$~0, which can be described by the continuous variation in contribution of the thermal accretion disk and non-thermal relativistic jet in this source. During the highest energy \g-ray jet flares, the OIR color variability is dominated by the internal dynamics of individual jet flares. From these long-term observations, we have developed a framework that can be tested for other blazars. We have undertaken a similar long-term OIR color variability study of the \smarts\ sample of blazars that will be reported in an upcoming paper (Isler et al., in preparation). Furthermore, we investigate the relationship between the OIR color variability and optical polarization in future papers.

\acknowledgements
\textit{Acknowledgements.} The authors wish to thank the anonymous referee for exceptionally useful comments that greatly improved the manuscript. J.C.I. receives support from the NSF Astronomy \& Astrophysics Postdoctoral Fellowship. \smarts\ observations of LAT-monitored blazars are supported by Yale University and \fermi\ GI grant NNX14AQ24G.  

\bibliography{Isler_ms_arXiv}

\begin{thebibliography}{}
\expandafter\ifx\csname natexlab\endcsname\relax\def\natexlab#1{#1}\fi

\bibitem[{{Abdo} {et~al.}(2009){Abdo}, {Ackermann}, {Ajello}, {Atwood},
  {Axelsson}, {Baldini}, {Ballet}, {Barbiellini}, {Bastieri}, {Baughman},
  {Bechtol}, {Bellazzini}, {Blandford}, {Bloom}, {Bonamente}, {Borgland},
  {Bouvier}, {Bregeon}, {Brez}, {Brigida}, {Bruel}, {Burnett}, {Caliandro},
  {Cameron}, {Caraveo}, {Casandjian}, {Cavazzuti}, {Cecchi}, {Charles},
  {Chekhtman}, {Chen}, {Cheung}, {Chiang}, {Ciprini}, {Claus}, {Cohen-Tanugi},
  {Colafrancesco}, {Collmar}, {Cominsky}, {Conrad}, {Costamante}, {Cutini},
  {Dermer}, {de Angelis}, {de Palma}, {Digel}, {do Couto e Silva}, {Drell},
  {Dubois}, {Dumora}, {Farnier}, {Favuzzi}, {Fegan}, {Ferrara}, {Finke},
  {Focke}, {Foschini}, {Frailis}, {Fuhrmann}, {Fukazawa}, {Funk}, {Fusco},
  {Gargano}, {Gasparrini}, {Gehrels}, {Germani}, {Giebels}, {Giglietto},
  {Giommi}, {Giordano}, {Giroletti}, {Glanzman}, {Godfrey}, {Grenier},
  {Grondin}, {Grove}, {Guillemot}, {Guiriec}, {Hanabata}, {Harding}, {Hartman},
  {Hayashida}, {Hays}, {Healey}, {Horan}, {Hughes}, {J{\'o}hannesson},
  {Johnson}, {Johnson}, {Johnson}, {Johnson}, {Kadler}, {Kamae}, {Katagiri},
  {Kataoka}, {Kerr}, {Kn{\"o}dlseder}, {Kocian}, {Kuehn}, {Kuss}, {Lande},
  {Latronico}, {Lemoine-Goumard}, {Longo}, {Loparco}, {Lott}, {Lovellette},
  {Lubrano}, {Madejski}, {Makeev}, {Massaro}, {Mazziotta}, {McConville},
  {McEnery}, {McGlynn}, {Meurer}, {Michelson}, {Mitthumsiri}, {Mizuno},
  {Moiseev}, {Monte}, {Monzani}, {Moretti}, {Morselli}, {Moskalenko}, {Murgia},
  {Nolan}, {Norris}, {Nuss}, {Ohsugi}, {Omodei}, {Orlando}, {Ormes}, {Ozaki},
  {Paneque}, {Panetta}, {Parent}, {Pelassa}, {Pepe}, {Pesce-Rollins}, {Piron},
  {Porter}, {Rain{\`o}}, {Rando}, {Razzano}, {Razzaque}, {Reimer}, {Reimer},
  {Reposeur}, {Reyes}, {Ritz}, {Rochester}, {Rodriguez}, {Romani}, {Ryde},
  {Sadrozinski}, {Sanchez}, {Sander}, {Saz Parkinson}, {Scargle}, {Schalk},
  {Sellerholm}, {Sgr{\`o}}, {Shaw}, {Smith}, {Smith}, {Spandre}, {Spinelli},
  {Starck}, {Strickman}, {Suson}, {Tajima}, {Takahashi}, {Takahashi}, {Tanaka},
  {Taylor}, {Thayer}, {Thayer}, {Thompson}, {Tibaldo}, {Torres}, {Tosti},
  {Tramacere}, {Uchiyama}, {Usher}, {Vilchez}, {Villata}, {Vitale}, {Waite},
  {Winer}, {Wood}, {Ylinen}, \& {Ziegler}}]{Abdo09}
{Abdo}, A.~A., {Ackermann}, M., {Ajello}, M., {et~al.} 2009, \apj, 700, 597

\bibitem[{{Abdo} {et~al.}(2010{\natexlab{a}}){Abdo}, {Ackermann}, {Ajello},
  {Axelsson}, {Baldini}, {Ballet}, {Barbiellini}, {Bastieri}, {Baughman},
  {Bechtol}, \& et~al.}]{Abdo10c}
---. 2010{\natexlab{a}}, \nat, 463, 919

\bibitem[{{Abdo} {et~al.}(2010{\natexlab{b}}){Abdo}, {Ackermann}, {Ajello},
  {Antolini}, {Baldini}, {Ballet}, {Barbiellini}, {Bastieri}, {Bechtol},
  {Bellazzini}, {Berenji}, {Blandford}, {Bloom}, {Bonamente}, {Borgland},
  {Bouvier}, {Bregeon}, {Brez}, {Brigida}, {Bruel}, {Buehler}, {Burnett},
  {Buson}, {Caliandro}, {Cameron}, {Caraveo}, {Carrigan}, {Casandjian},
  {Cavazzuti}, {Cecchi}, {{C}elik}, {Chekhtman}, {Cheung}, {Chiang}, {Ciprini},
  {Claus}, {Cohen-Tanugi}, {Cominsky}, {Conrad}, {Costamante}, {Cutini},
  {Dermer}, {de Angelis}, {de Palma}, {Silva}, {Drell}, {Dubois}, {Dumora},
  {Farnier}, {Favuzzi}, {Fegan}, {Focke}, {Fortin}, {Frailis}, {Fukazawa},
  {Funk}, {Fusco}, {Gargano}, {Gasparrini}, {Gehrels}, {Germani}, {Giebels},
  {Giglietto}, {Giommi}, {Giordano}, {Glanzman}, {Godfrey}, {Grenier},
  {Grondin}, {Grove}, {Guiriec}, {Hadasch}, {Hayashida}, {Hays}, {Healey},
  {Horan}, {Hughes}, {Itoh}, {J{\'o}hannesson}, {Johnson}, {Johnson}, {Kamae},
  {Katagiri}, {Kataoka}, {Kawai}, {Kn{\"o}dlseder}, {Kuss}, {Lande}, {Larsson},
  {Latronico}, {Lemoine-Goumard}, {Longo}, {Loparco}, {Lott}, {Lovellette},
  {Lubrano}, {Madejski}, {Makeev}, {Massaro}, {Mazziotta}, {McEnery},
  {Michelson}, {Mitthumsiri}, {Mizuno}, {Moiseev}, {Monte}, {Monzani},
  {Morselli}, {Moskalenko}, {Mueller}, {Murgia}, {Nolan}, {Norris}, {Nuss},
  {Ohno}, {Ohsugi}, {Omodei}, {Orlando}, {Ormes}, {Ozaki}, {Panetta}, {Parent},
  {Pelassa}, {Pepe}, {Pesce-Rollins}, {Piron}, {Porter}, {Rain{\`o}}, {Rando},
  {Razzano}, {Reimer}, {Reimer}, {Ritz}, {Rodriguez}, {Romani}, {Roth}, {Ryde},
  {Sadrozinski}, {Sander}, {Scargle}, {Sgr{\`o}}, {Shaw}, {Smith}, {Spandre},
  {Spinelli}, {Starck}, {Strickman}, {Suson}, {Takahashi}, {Takahashi},
  {Tanaka}, {Thayer}, {Thayer}, {Thompson}, {Tibaldo}, {Torres}, {Tosti},
  {Tramacere}, {Uchiyama}, {Usher}, {Vasileiou}, {Vilchez}, {Vitale}, {Waite},
  {Wallace}, {Wang}, {Winer}, {Wood}, {Yang}, {Ylinen}, \& {Ziegler}}]{Abdo10}
---. 2010{\natexlab{b}}, \apj, 722, 520

\bibitem[{{Abraham} \& {Carrara}(1998)}]{Abraham98}
{Abraham}, Z., \& {Carrara}, E.~A. 1998, \apj, 496, 172

\bibitem[{{Antonucci}(1993)}]{antonucci93}
{Antonucci}, R. 1993, \araa, 31, 473

\bibitem[{{Bonning} {et~al.}(2012){Bonning}, {Urry}, {Bailyn}, {Buxton},
  {Chatterjee}, {Coppi}, {Fossati}, {Isler}, \& {Maraschi}}]{Bonning12}
{Bonning}, E., {Urry}, C.~M., {Bailyn}, C., {et~al.} 2012, \apj, 756, 13

\bibitem[{{B{\"o}ttcher} {et~al.}(2005){B{\"o}ttcher}, {Harvey}, {Joshi},
  {Villata}, {Raiteri}, {Bramel}, {Mukherjee}, {Savolainen}, {Cui}, {Fossati},
  {Smith}, {Able}, {Aller}, {Aller}, {Arkharov}, {Augusteijn}, {Baliyan},
  {Barnaby}, {Berdyugin}, {Ben{\'{\i}}tez}, {Boltwood}, {Carini}, {Carosati},
  {Ciprini}, {Coloma}, {Crapanzano}, {de Diego}, {Di Paola}, {Dolci}, {Fan},
  {Frasca}, {Hagen-Thorn}, {Horan}, {Ibrahimov}, {Kimeridze}, {Kovalev},
  {Kovalev}, {Kurtanidze}, {L{\"a}hteenm{\"a}ki}, {Lanteri}, {Larionov},
  {Larionova}, {Lindfors}, {Marilli}, {Mirabal}, {Nikolashvili}, {Nilsson},
  {Ohlert}, {Ohnishi}, {Oksanen}, {Ostorero}, {Oyer}, {Papadakis}, {Pasanen},
  {Poteet}, {Pursimo}, {Sadakane}, {Sigua}, {Takalo}, {Tartar},
  {Ter{\"a}sranta}, {Tosti}, {Walters}, {Wiik}, {Wilking}, {Wills}, {Xilouris},
  {Fletcher}, {Gu}, {Lee}, {Pak}, \& {Yim}}]{Bottcher05}
{B{\"o}ttcher}, M., {Harvey}, J., {Joshi}, M., {et~al.} 2005, \apj, 631, 169

\bibitem[{{Caproni} {et~al.}(2013){Caproni}, {Abraham}, \&
  {Monteiro}}]{Caproni13}
{Caproni}, A., {Abraham}, Z., \& {Monteiro}, H. 2013, \mnras, 428, 280

\bibitem[{{Chatterjee} {et~al.}(2012){Chatterjee}, {Bailyn}, {Bonning},
  {Buxton}, {Coppi}, {Fossati}, {Isler}, {Maraschi}, \& {Urry}}]{Chatterjee12}
{Chatterjee}, R., {Bailyn}, C.~D., {Bonning}, E.~W., {et~al.} 2012, \apj, 749,
  191

\bibitem[{{Chatterjee} {et~al.}(2013){Chatterjee}, {Fossati}, {Urry}, {Bailyn},
  {Maraschi}, {Buxton}, {Bonning}, {Isler}, \& {Coppi}}]{Chatterjee13}
{Chatterjee}, R., {Fossati}, G., {Urry}, C.~M., {et~al.} 2013, \apjl, 763, L11

\bibitem[{{D'Amicis} {et~al.}(2002){D'Amicis}, {Nesci}, {Massaro}, {Maesano},
  {Montagni}, \& {D'Alessio}}]{Damicis02}
{D'Amicis}, R., {Nesci}, R., {Massaro}, E., {et~al.} 2002, \pasa, 19, 111

\bibitem[{{Fossati} {et~al.}(1998){Fossati}, {Maraschi}, {Celotti}, {Comastri},
  \& {Ghisellini}}]{Fossati98}
{Fossati}, G., {Maraschi}, L., {Celotti}, A., {Comastri}, A., \& {Ghisellini},
  G. 1998, \mnras, 299, 433

\bibitem[{{Frank} {et~al.}(2002){Frank}, {King}, \& {Raine}}]{Frank02}
{Frank}, J., {King}, A., \& {Raine}, D.~J. 2002, {Accretion Power in
  Astrophysics: Third Edition, Cambridge University Press}

\bibitem[{{Gaur} {et~al.}(2012){Gaur}, {Gupta}, {Strigachev}, {Bachev},
  {Semkov}, {Wiita}, {Peneva}, {Boeva}, {Slavcheva-Mihova}, {Mihov}, {Latev},
  \& {Pandey}}]{Gaur12}
{Gaur}, H., {Gupta}, A.~C., {Strigachev}, A., {et~al.} 2012, \mnras, 425, 3002

\bibitem[{{Ghisellini}(2013)}]{Ghisellini13b}
{Ghisellini}, G. 2013, in European Physical Journal Web of Conferences,
  Vol.~61, European Physical Journal Web of Conferences, 5001

\bibitem[{{Ghisellini} {et~al.}(1997){Ghisellini}, {Villata}, {Raiteri},
  {Bosio}, {de Francesco}, {Latini}, {Maesano}, {Massaro}, {Montagni}, {Nesci},
  {Tosti}, {Fiorucci}, {Pian}, {Maraschi}, {Treves}, {Comastri}, \&
  {Mignoli}}]{Ghisellini97}
{Ghisellini}, G., {Villata}, M., {Raiteri}, C.~M., {et~al.} 1997, \aap, 327, 61

\bibitem[{{Ghosh} {et~al.}(2000){Ghosh}, {Ramsey}, {Sadun}, \&
  {Soundararajaperumal}}]{Ghosh00}
{Ghosh}, K.~K., {Ramsey}, B.~D., {Sadun}, A.~C., \& {Soundararajaperumal}, S.
  2000, \apjs, 127, 11

\bibitem[{{Gower} {et~al.}(1982){Gower}, {Gregory}, {Unruh}, \&
  {Hutchings}}]{Gower82}
{Gower}, A.~C., {Gregory}, P.~C., {Unruh}, W.~G., \& {Hutchings}, J.~B. 1982,
  \apj, 262, 478

\bibitem[{{Gu} \& {Ai}(2011{\natexlab{a}})}]{Gu11b}
{Gu}, M., \& {Ai}, Y.~L. 2011{\natexlab{a}}, Journal of Astrophysics and
  Astronomy, 32, 87

\bibitem[{{Gu} \& {Ai}(2011{\natexlab{b}})}]{Gu11}
{Gu}, M.-F., \& {Ai}, Y.~L. 2011{\natexlab{b}}, \aap, 528, A95

\bibitem[{{Gu} {et~al.}(2006){Gu}, {Lee}, {Pak}, {Yim}, \& {Fletcher}}]{Gu06}
{Gu}, M.~F., {Lee}, C.-U., {Pak}, S., {Yim}, H.~S., \& {Fletcher}, A.~B. 2006,
  \aap, 450, 39

\bibitem[{{Hartman} {et~al.}(1996){Hartman}, {Webb}, {Marscher}, {Travis},
  {Dermer}, {Aller}, {Aller}, {Balonek}, {Bennett}, {Bloom}, {Fujimoto},
  {Hermsen}, {Hughes}, {Jenkins}, {Kii}, {Kurfess}, {Makino}, {Mattox}, {von
  Montigny}, {Ohashi}, {Robson}, {Ryan}, {Sadun}, {Schoenfelder}, {Smith},
  {Teraesranta}, {Tornikoski}, \& {Turner}}]{Hartman96}
{Hartman}, R.~C., {Webb}, J.~R., {Marscher}, A.~P., {et~al.} 1996, \apj, 461,
  698

\bibitem[{{Hovatta}(2017)}]{Hovatta17}
{Hovatta}, T. 2017, ArXiv e-prints, arXiv:1701.07997

\bibitem[{{Kiehlmann} {et~al.}(2016){Kiehlmann}, {Savolainen}, {Jorstad},
  {Sokolovsky}, {Schinzel}, {Marscher}, {Larionov}, {Agudo}, {Akitaya},
  {Ben{\'{\i}}tez}, {Berdyugin}, {Blinov}, {Bochkarev}, {Borman}, {Burenkov},
  {Casadio}, {Doroshenko}, {Efimova}, {Fukazawa}, {G{\'o}mez}, {Grishina},
  {Hagen-Thorn}, {Heidt}, {Hiriart}, {Itoh}, {Joshi}, {Kawabata}, {Kimeridze},
  {Kopatskaya}, {Korobtsev}, {Krajci}, {Kurtanidze}, {Kurtanidze}, {Larionova},
  {Larionova}, {Lindfors}, {L{\'o}pez}, {McHardy}, {Molina}, {Moritani},
  {Morozova}, {Nazarov}, {Nikolashvili}, {Nilsson}, {Pulatova}, {Reinthal},
  {Sadun}, {Sasada}, {Savchenko}, {Sergeev}, {Sigua}, {Smith}, {Sorcia},
  {Spiridonova}, {Takaki}, {Takalo}, {Taylor}, {Troitsky}, {Uemura},
  {Ugolkova}, {Ui}, {Yoshida}, {Zensus}, \& {Zhdanova}}]{Kiehlmann16}
{Kiehlmann}, S., {Savolainen}, T., {Jorstad}, S.~G., {et~al.} 2016, \aap, 590,
  A10

\bibitem[{{Larionov} {et~al.}(2008){Larionov}, {Jorstad}, {Marscher},
  {Raiteri}, {Villata}, {Agudo}, {Aller}, {Arkharov}, {Asfandiyarov}, {Bach},
  {Bachev}, {Berdyugin}, {B{\"o}ttcher}, {Buemi}, {Calcidese}, {Carosati},
  {Charlot}, {Chen}, {di Paola}, {Dolci}, {Dogru}, {Doroshenko}, {Efimov},
  {Erdem}, {Frasca}, {Fuhrmann}, {Giommi}, {Glowienka}, {Gupta}, {Gurwell},
  {Hagen-Thorn}, {Hsiao}, {Ibrahimov}, {Jordan}, {Kamada}, {Konstantinova},
  {Kopatskaya}, {Kovalev}, {Kovalev}, {Kurtanidze}, {L{\"a}hteenm{\"a}ki},
  {Lanteri}, {Larionova}, {Leto}, {Le Campion}, {Lee}, {Lindfors}, {Marilli},
  {McHardy}, {Mingaliev}, {Nazarov}, {Nieppola}, {Nilsson}, {Ohlert},
  {Pasanen}, {Porter}, {Pursimo}, {Ros}, {Sadakane}, {Sadun}, {Sergeev},
  {Smith}, {Strigachev}, {Sumitomo}, {Takalo}, {Tanaka}, {Trigilio}, {Umana},
  {Ungerechts}, {Volvach}, \& {Yuan}}]{Larionov08}
{Larionov}, V.~M., {Jorstad}, S.~G., {Marscher}, A.~P., {et~al.} 2008, \aap,
  492, 389

\bibitem[{{Lira} {et~al.}(2011){Lira}, {Ar{\'e}valo}, {Uttley}, {McHardy}, \&
  {Breedt}}]{Lira11}
{Lira}, P., {Ar{\'e}valo}, P., {Uttley}, P., {McHardy}, I., \& {Breedt}, E.
  2011, \mnras, 415, 1290

\bibitem[{{Maraschi} {et~al.}(1994){Maraschi}, {Grandi}, {Urry}, {Wehrle},
  {Madejski}, {Fink}, {Ghisellini}, {Hartman}, {Koratkar}, {von Montigny},
  {Pian}, {Thomas}, {Treves}, {Aller}, {Aller}, {Bailyn}, {Balonek}, {Bock},
  {Collmar}, {Glass}, {Litchfield}, {McHardy}, {Mendez}, {Pesce}, {Reuter},
  {Robson}, {Steppe}, {Stevens}, {Teraesranta}, \& {Wagner}}]{Maraschi94}
{Maraschi}, L., {Grandi}, P., {Urry}, C.~M., {et~al.} 1994, \apjl, 435, L91

\bibitem[{{Massaro} {et~al.}(1998){Massaro}, {Nesci}, {Maesano}, {Montagni}, \&
  {D'Alessio}}]{Massaro98}
{Massaro}, E., {Nesci}, R., {Maesano}, M., {Montagni}, F., \& {D'Alessio}, F.
  1998, \mnras, 299, 47

\bibitem[{{Mattox} {et~al.}(1996){Mattox}, {Bertsch}, {Chiang}, {Dingus},
  {Digel}, {Esposito}, {Fierro}, {Hartman}, {Hunter}, {Kanbach}, {Kniffen},
  {Lin}, {Macomb}, {Mayer-Hasselwander}, {Michelson}, {von Montigny},
  {Mukherjee}, {Nolan}, {Ramanamurthy}, {Schneid}, {Sreekumar}, {Thompson}, \&
  {Willis}}]{Mattox96}
{Mattox}, J.~R., {Bertsch}, D.~L., {Chiang}, J., {et~al.} 1996, \apj, 461, 396

\bibitem[{{Meyer} {et~al.}(2012){Meyer}, {Fossati}, {Georganopoulos}, \&
  {Lister}}]{Meyer12b}
{Meyer}, E.~T., {Fossati}, G., {Georganopoulos}, M., \& {Lister}, M.~L. 2012,
  ArXiv e-prints, arXiv:1205.0794

\bibitem[{{Monet} {et~al.}(2003){Monet}, {Levine}, {Canzian}, {Ables}, {Bird},
  {Dahn}, {Guetter}, {Harris}, {Henden}, {Leggett}, {Levison}, {Luginbuhl},
  {Martini}, {Monet}, {Munn}, {Pier}, {Rhodes}, {Riepe}, {Sell}, {Stone},
  {Vrba}, {Walker}, {Westerhout}, {Brucato}, {Reid}, {Schoening}, {Hartley},
  {Read}, \& {Tritton}}]{Monet03}
{Monet}, D.~G., {Levine}, S.~E., {Canzian}, B., {et~al.} 2003, \aj, 125, 984

\bibitem[{{Osterman Meyer} {et~al.}(2009){Osterman Meyer}, {Miller},
  {Marshall}, {Ryle}, {Aller}, {Aller}, \& {Balonek}}]{Osterman09}
{Osterman Meyer}, A., {Miller}, H.~R., {Marshall}, K., {et~al.} 2009, \aj, 138,
  1902

\bibitem[{{Osterman Meyer} {et~al.}(2008){Osterman Meyer}, {Miller},
  {Marshall}, {Ryle}, {Aller}, {Aller}, {McFarland}, {Pollock}, {Reichart},
  {Crain}, {Ivarsen}, {La Cluyze}, \& {Nysewander}}]{Osterman08}
---. 2008, \aj, 136, 1398

\bibitem[{{Padovani} \& {Giommi}(1996)}]{Padovani96}
{Padovani}, P., \& {Giommi}, P. 1996, \mnras, 279, 526

\bibitem[{{Pian} {et~al.}(2005){Pian}, {Falomo}, \& {Treves}}]{Pian05}
{Pian}, E., {Falomo}, R., \& {Treves}, A. 2005, \mnras, 361, 919

\bibitem[{{Pian} {et~al.}(1999){Pian}, {Urry}, {Maraschi}, {Madejski},
  {McHardy}, {Koratkar}, {Treves}, {Chiappetti}, {Grandi}, {Hartman}, {Kubo},
  {Leach}, {Pesce}, {Imhoff}, {Thompson}, \& {Wehrle}}]{Pian99}
{Pian}, E., {Urry}, C.~M., {Maraschi}, L., {et~al.} 1999, \apj, 521, 112

\bibitem[{{Qian} {et~al.}(2014){Qian}, {Britzen}, {Witzel}, {Krichbaum}, {Gan},
  \& {Gao}}]{Qian14}
{Qian}, S.-J., {Britzen}, S., {Witzel}, A., {et~al.} 2014, Research in
  Astronomy and Astrophysics, 14, 249

\bibitem[{{Ram{\'{\i}}rez} {et~al.}(2004){Ram{\'{\i}}rez}, {de Diego},
  {Dultzin-Hacyan}, \& {Gonz{\'a}lez-P{\'e}rez}}]{Ramirez04}
{Ram{\'{\i}}rez}, A., {de Diego}, J.~A., {Dultzin-Hacyan}, D., \&
  {Gonz{\'a}lez-P{\'e}rez}, J.~N. 2004, \aap, 421, 83

\bibitem[{{Ruan} {et~al.}(2014){Ruan}, {Anderson}, {Dexter}, \&
  {Agol}}]{Ruan14}
{Ruan}, J.~J., {Anderson}, S.~F., {Dexter}, J., \& {Agol}, E. 2014, ArXiv
  e-prints, arXiv:1401.1211

\bibitem[{{Shakura} \& {Sunyaev}(1973)}]{Shakura73}
{Shakura}, N.~I., \& {Sunyaev}, R.~A. 1973, \aap, 24, 337

\bibitem[{{Skrutskie} {et~al.}(2006){Skrutskie}, {Cutri}, {Stiening},
  {Weinberg}, {Schneider}, {Carpenter}, {Beichman}, {Capps}, {Chester},
  {Elias}, {Huchra}, {Liebert}, {Lonsdale}, {Monet}, {Price}, {Seitzer},
  {Jarrett}, {Kirkpatrick}, {Gizis}, {Howard}, {Evans}, {Fowler}, {Fullmer},
  {Hurt}, {Light}, {Kopan}, {Marsh}, {McCallon}, {Tam}, {Van Dyk}, \&
  {Wheelock}}]{Skrutskie06}
{Skrutskie}, M.~F., {Cutri}, R.~M., {Stiening}, R., {et~al.} 2006, \aj, 131,
  1163

\bibitem[{{Stalin} {et~al.}(2006){Stalin}, {Gopal-Krishna}, {Sagar}, {Wiita},
  {Mohan}, \& {Pandey}}]{Stalin06}
{Stalin}, C.~S., {Gopal-Krishna}, {Sagar}, R., {et~al.} 2006, \mnras, 366, 1337

\bibitem[{{Tr{\`e}vese} \& {Vagnetti}(2002)}]{Trevese02}
{Tr{\`e}vese}, D., \& {Vagnetti}, F. 2002, \apj, 564, 624

\bibitem[{{Urry} \& {Padovani}(1995)}]{Urry95}
{Urry}, C.~M., \& {Padovani}, P. 1995, \pasp, 107, 803

\bibitem[{{Vagnetti} {et~al.}(2003){Vagnetti}, {Trevese}, \&
  {Nesci}}]{Vagnetti03}
{Vagnetti}, F., {Trevese}, D., \& {Nesci}, R. 2003, \apj, 590, 123

\bibitem[{{Villata} {et~al.}(2002){Villata}, {Raiteri}, {Kurtanidze},
  {Nikolashvili}, {Ibrahimov}, {Papadakis}, {Tsinganos}, {Sadakane}, {Okada},
  {Takalo}, {Sillanp{\"a}{\"a}}, {Tosti}, {Ciprini}, {Frasca}, {Marilli},
  {Robb}, {Noble}, {Jorstad}, {Hagen-Thorn}, {Larionov}, {Nesci}, {Maesano},
  {Schwartz}, {Basler}, {Gorham}, {Iwamatsu}, {Kato}, {Pullen},
  {Ben{\'{\i}}tez}, {de Diego}, {Moilanen}, {Oksanen}, {Rodriguez}, {Sadun},
  {Kelly}, {Carini}, {Miller}, {Catalano}, {Dultzin-Hacyan}, {Fan}, {Ishioka},
  {Karttunen}, {Kein{\"a}nen}, {Kudryavtseva}, {Lainela}, {Lanteri},
  {Larionova}, {Matsumoto}, {Mattox}, {Montagni}, {Nucciarelli}, {Ostorero},
  {Papamastorakis}, {Pasanen}, {Sobrito}, \& {Uemura}}]{Villata02}
{Villata}, M., {Raiteri}, C.~M., {Kurtanidze}, O.~M., {et~al.} 2002, \aap, 390,
  407

\bibitem[{{Villata} {et~al.}(2006){Villata}, {Raiteri}, {Balonek}, {Aller},
  {Jorstad}, {Kurtanidze}, {Nicastro}, {Nilsson}, {Aller}, {Arai}, {Arkharov},
  {Bach}, {Ben{\'{\i}}tez}, {Berdyugin}, {Buemi}, {B{\"o}ttcher}, {Carosati},
  {Casas}, {Caulet}, {Chen}, {Chiang}, {Chou}, {Ciprini}, {Coloma}, {di Rico},
  {D{\'{\i}}az}, {Efimova}, {Forsyth}, {Frasca}, {Fuhrmann}, {Gadway}, {Gupta},
  {Hagen-Thorn}, {Harvey}, {Heidt}, {Hernandez-Toledo}, {Hroch}, {Hu}, {Hudec},
  {Ibrahimov}, {Imada}, {Kamata}, {Kato}, {Katsuura}, {Konstantinova},
  {Kopatskaya}, {Kotaka}, {Kovalev}, {Kovalev}, {Krichbaum}, {Kubota},
  {Kurosaki}, {Lanteri}, {Larionov}, {Larionova}, {Laurikainen}, {Lee}, {Leto},
  {L{\"a}hteenm{\"a}ki}, {L{\'o}pez-Cruz}, {Marilli}, {Marscher}, {McHardy},
  {Mondal}, {Mullan}, {Napoleone}, {Nikolashvili}, {Ohlert}, {Postnikov},
  {Pursimo}, {Ragni}, {Ros}, {Sadakane}, {Sadun}, {Savolainen}, {Sergeeva},
  {Sigua}, {Sillanp{\"a}{\"a}}, {Sixtova}, {Sumitomo}, {Takalo},
  {Ter{\"a}sranta}, {Tornikoski}, {Trigilio}, {Umana}, {Volvach}, {Voss}, \&
  {Wortel}}]{Villata06}
{Villata}, M., {Raiteri}, C.~M., {Balonek}, T.~J., {et~al.} 2006, \aap, 453,
  817

\bibitem[{{Wehrle} {et~al.}(1998){Wehrle}, {Pian}, {Urry}, {Maraschi},
  {McHardy}, {Lawson}, {Ghisellini}, {Hartman}, {Madejski}, {Makino},
  {Marscher}, {Wagner}, {Webb}, {Aldering}, {Aller}, {Aller}, {Backman},
  {Balonek}, {Boltwood}, {Bonnell}, {Caplinger}, {Celotti}, {Collmar},
  {Dalton}, {Drucker}, {Falomo}, {Fichtel}, {Freudling}, {Gear},
  {Gonzalez-Perez}, {Hall}, {Inoue}, {Johnson}, {Kazanas}, {Kidger}, {Kii},
  {Kollgaard}, {Kondo}, {Kurfess}, {Lin}, {McCollum}, {McNaron-Brown},
  {Nagase}, {Nair}, {Penton}, {Pesce}, {Pohl}, {Raiteri}, {Renda}, {Robson},
  {Sambruna}, {Schirmer}, {Shrader}, {Sikora}, {Sillanpaeae}, {Smith},
  {Stevens}, {Stocke}, {Takalo}, {Teraesranta}, {Thompson}, {Thompson},
  {Tornikoski}, {Tosti}, {Treves}, {Turcotte}, {Unwin}, {Valtaoja}, {Villata},
  {Xu}, {Yamashita}, \& {Zook}}]{Wehrle98}
{Wehrle}, A.~E., {Pian}, E., {Urry}, C.~M., {et~al.} 1998, \apj, 497, 178

\bibitem[{{Wu} {et~al.}(2011){Wu}, {Zhou}, {Ma}, \& {Jiang}}]{Wu11}
{Wu}, J., {Zhou}, X., {Ma}, J., \& {Jiang}, Z. 2011, \mnras, 418, 1640

\bibitem[{{Zhang} {et~al.}(2015){Zhang}, {Chen}, {B{\"o}ttcher}, {Guo}, \&
  {Li}}]{Zhang15}
{Zhang}, H., {Chen}, X., {B{\"o}ttcher}, M., {Guo}, F., \& {Li}, H. 2015, \apj,
  804, 58

\end{thebibliography}

\end{document}